\documentclass[11pt,preprintnumbers,titlepage]{revtex4-2}
\usepackage[latin9]{inputenc}
\usepackage{amsmath}
\usepackage{amssymb}
\usepackage{graphicx}
\usepackage[unicode=true,
 bookmarks=false,
 breaklinks=false,pdfborder={0 0 1},backref=section,colorlinks=false]
 {hyperref}
\hypersetup{
 colorlinks,linkcolor=blue,citecolor=blue,urlcolor=blue}

\makeatletter

\usepackage{amsfonts}
\usepackage{subfigure}
\usepackage{float}
\usepackage{cleveref}
\usepackage{listings}
\usepackage{xcolor}
\usepackage{array}
\usepackage{url}
\usepackage{color}

\@ifundefined{textcolor}{}{
	\definecolor{BLACK}{gray}{0}
	\definecolor{WHITE}{gray}{1}
	\definecolor{RED}{rgb}{1,0,0}
	\definecolor{GREEN}{rgb}{0,1,0}
	\definecolor{BLUE}{rgb}{0,0,1}
	\definecolor{CYAN}{cmyk}{1,0,0,0}
	\definecolor{MAGENTA}{cmyk}{0,1,0,0}
	\definecolor{YELLOW}{cmyk}{0,0,1,0}
}

\makeatother

\begin{document}
\preprint{CTP-SCU/2022007}
\title{Gravitational Lensing by Black Holes with Multiple Photon Spheres}
\author{Guangzhou Guo$^{a}$}
\email{gzguo@stu.scu.edu.cn}

\author{Xin Jiang$^{a}$}
\email{xjang@stu.scu.edu.cn}

\author{Peng Wang$^{a}$}
\email{pengw@scu.edu.cn}

\author{Houwen Wu$^{a,b}$}
\email{hw598@damtp.cam.ac.uk}

\affiliation{$^{a}$Center for Theoretical Physics, College of Physics, Sichuan
University, Chengdu, 610064, China}
\affiliation{$^{b}$Department of Applied Mathematics and Theoretical Physics,
University of Cambridge, Wilberforce Road, Cambridge, CB3 0WA, UK}
\begin{abstract}
We study gravitational lensing of light by hairy black holes, which,
in a certain parameter regime, can possess two photon spheres of different
size outside the event horizon. In particular, we focus on higher-order
images of a point-like light source and a\ luminous celestial sphere
produced by strong gravitational lensing near photon spheres. Two
photon spheres usually triple the number of high-order images of a
point-like light source. When a hairy black hole is illuminated by
a celestial sphere, two photon spheres would give rise to two critical
curves in the black hole image, and the smaller critical curve coincides
with the shadow edge. In addition to a set of higher-order images
of the celestial sphere outside the shadow edge, two more sets of
higher-order images are observed inside and outside the larger critical
curve, respectively.
\end{abstract}
\maketitle
\tableofcontents{}

\section{Introduction}

\label{sec:Introduction}

One of the elegant predictions of general relativity is the bending
of light rays in curved space, which produces an effect similar to
that of a lens, and hence is known as gravitational lensing \cite{Dyson:1920cwa,Einstein:1936llh,Eddington:1987tk}.
Gravitational lensing has long been an essential tool to address fundamental
problems in astrophysics and cosmology. The lensing near relativistic
bodies, including a star on an orbit in the Kerr spacetime \cite{Cunningham:1972opt}
and an accretion disk around a Schwarzschild black hole \cite{Luminet:1979nyg},
were first studied in 1970s. Since then, there have been many works
investigating gravitational lensing by the distribution of structures
\cite{Mellier:1998pk,Bartelmann:1999yn,Heymans:2013fya}, dark matter
\cite{Kaiser:1992ps,Clowe:2006eq,Atamurotov:2021hoq}, dark energy
\cite{Biesiada:2006zf,Cao:2015qja,DES:2020ahh,DES:2021wwk,Zhang:2021ygh},
quasars \cite{SDSS:2000jpb,Peng:2006ew,Oguri:2010ns,Yue:2021nwt},
gravitational waves \cite{Seljak:2003pn,Diego:2021fyd,Finke:2021znb}
and some other compact objects \cite{Schmidt:2008hc,Guzik:2009cm,Liao:2015uzb,Goulart:2017iko,Nascimento:2020ime,Junior:2021svb,Islam:2021ful,Tsukamoto:2022vkt}.
Recently, the Event Horizon Telescope collaboration achieved an angular
resolution sufficient to observe the image of the supermassive black
hole in the center of galaxy M87, which ushered us into a new era
of studying gravitational lensing in the strong gravity regime \cite{Akiyama:2019cqa,Akiyama:2019brx,Akiyama:2019sww,Akiyama:2019bqs,Akiyama:2019fyp,Akiyama:2019eap,Akiyama:2021qum,Akiyama:2021tfw}.
In particular, the shadow in black hole images, which is closely related
to strong gravitational lensing near photon spheres, has been a subject
of great interest to researchers \cite{Falcke:1999pj,Virbhadra:1999nm,Claudel:2000yi,Eiroa:2002mk,Virbhadra:2008ws,Yumoto:2012kz,Wei:2013kza,Zakharov:2014lqa,Atamurotov:2015xfa,Cunha:2016wzk,Dastan:2016bfy,Amir:2017slq,Wang:2017hjl,Ovgun:2018tua,Perlick:2018iye,Kumar:2019pjp,Zhu:2019ura,Ma:2019ybz,Mishra:2019trb,Zeng:2020dco,Zeng:2020vsj,Qin:2020xzu,Saurabh:2020zqg,Roy:2020dyy,Li:2020drn,Kumar:2020hgm,Zhang:2020xub,Olmo:2021piq,Guerrero:2022qkh,Virbhadra:2022iiy}.

To understand the formation of Hairy Black Holes (HBHs), a novel type
of HBH solutions have recently been constructed in Einstein-Maxwell-scalar
(EMS) models, where the non-minimal coupling between the scalar field
and the electromagnetic field can trigger a tachyonic instability
to form spontaneously scalarized HBHs from Reissner-Nordström (RN)
black holes \cite{Herdeiro:2018wub,Konoplya:2019goy,Wang:2020ohb,Guo:2021zed,Guo:2021ere}.
Properties of the HBHs have been extensively studied in the literature,
e.g., different non-minimal coupling functions \cite{Fernandes:2019rez,Fernandes:2019kmh,Blazquez-Salcedo:2020nhs},
massive and self-interacting scalar fields \cite{Zou:2019bpt,Fernandes:2020gay},
horizonless reflecting stars \cite{Peng:2019cmm}, stability analysis
of HBHs \cite{Myung:2018vug,Myung:2019oua,Zou:2020zxq,Myung:2020etf,Mai:2020sac},
higher dimensional scalar-tensor models \cite{Astefanesei:2020qxk},
quasinormal modes of HBHs \cite{Myung:2018jvi,Blazquez-Salcedo:2020jee},
two U$\left(1\right)$ fields \cite{Myung:2020dqt}, quasi-topological
electromagnetism \cite{Myung:2020ctt}, topology and spacetime structure
influences \cite{Guo:2020zqm}, and HBHs in the dS/AdS spacetime \cite{Brihaye:2019dck,Brihaye:2019gla,Zhang:2021etr,Guo:2021zed}.

Intriguingly, the scalarized HBHs have been found to possess two photon
spheres outside the event horizon in certain parameter regions \cite{Gan:2021pwu,Gan:2021xdl}.
The existence of two photon spheres can significantly affect the optical
appearance of HBHs illuminated by the surrounding accretion disk,
e.g., producing bright rings of different radus in the black hole
images \cite{Gan:2021pwu} and noticeably increasing the flux of the
observed images \cite{Gan:2021xdl}. Moreover, the effective potential
for a scalar perturbation in the HBHs with two photon spheres was
shown to exhibit a double-peak structure, leading to long-lived quasinormal
modes \cite{Guo:2021enm} and echo signals \cite{Guo:2022umh}. It
is worth noting that the existence of two photon spheres outside the
event horizon has also been reported for dyonic black holes with a
quasi-topological electromagnetic term \cite{Liu:2019rib} and black
holes in massive gravity \cite{deRham:2010kj,Dong:2020odp}.

Since gravitational lensing plays a key role in observing black holes,
we aim to study gravitational lensing by the HBHs with two photon
spheres in this paper. The remainder of this paper is organized as
follows. After we briefly review the HBH solutions in the Einstein-Maxwell-scalar
theory and introduce the observational settings in section \ref{sec:setup},
the lensed images of a point-like light source and a\ celestial sphere
are presented and discussed in section \ref{sec:Numresult}. Section
\ref{sec:CON} is devoted to our conclusions. We set $G=c=1$ throughout
the paper.

\section{Set up}

\label{sec:setup} 
\begin{figure}[tb]
\includegraphics[width=1\textwidth]{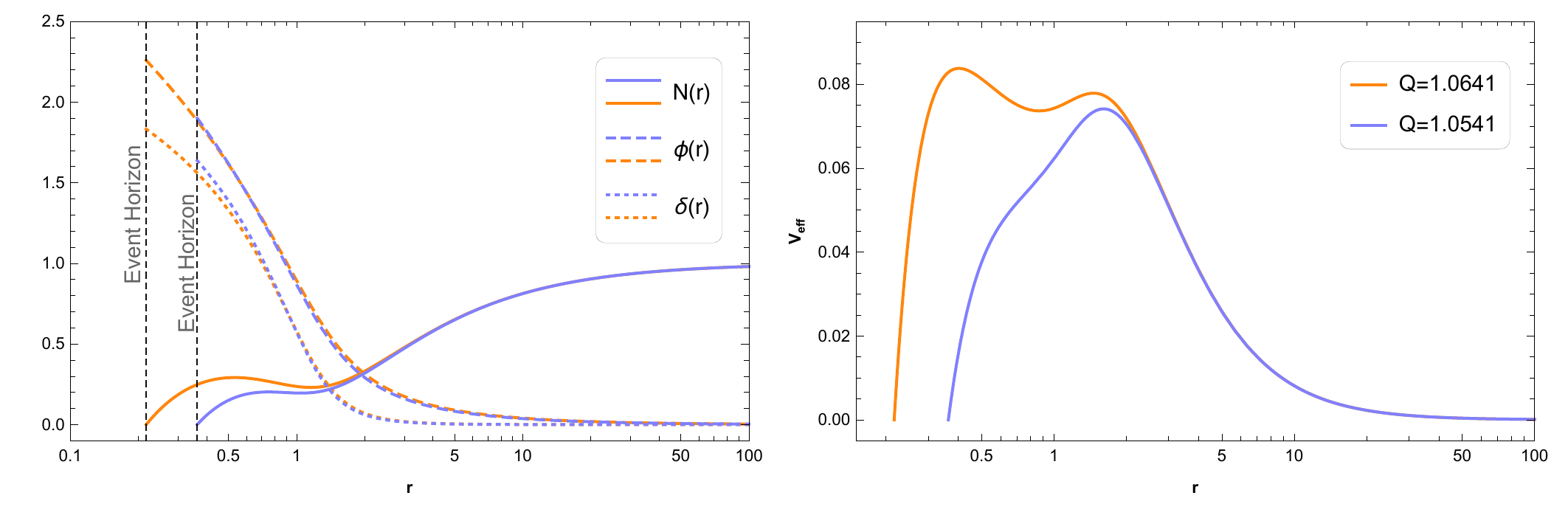}\caption{HBH solutions with $a=0.9$ and $M=1$ for $Q=1.0541$ (blue lines)
and $Q=1.0641$ (orange lines), and the associated effective potentials
as a function of $r$. \textbf{Left}: The metric functions are plotted
outside the event horizon (vertical dashed lines), and the solid,
dashed and dotted lines represent $N\left(r\right)$, $\phi\left(r\right)$
and $\delta\left(r\right)$, respectively. \textbf{Right}: For the
smaller charge (blue line), the effective potential has only a single
extremum. For the larger charge (orange line), the effective potential
presents a double-peak structure with two local maxima, corresponding
to two photon spheres outside the event horizon.}
\label{fig:HBH}
\end{figure}

In this section, we briefly review the HBH solutions, discuss the
associated geodesic equations and introduce the observational model.
Consider an Einstein-Maxwell-scalar theory with an exponential coupling,
whose action is given by \cite{Herdeiro:2018wub} 
\begin{equation}
S=\int d^{4}x\sqrt{-g}\left[\mathcal{R}-2\partial_{\mu}\phi\partial^{\mu}\phi-e^{a\phi^{2}}F_{\mu\nu}F^{\mu\nu}\right],\label{eq:metric}
\end{equation}
where $\mathcal{R}$ is the Ricci scalar, the scalar field $\phi$
is minimally coupled to the metric $g_{\mu\nu}$ and non-minimally
coupled to the electromagnetic field $A_{\mu}$, and $F_{\mu\nu}=\partial_{\mu}A_{\nu}-\partial_{\nu}A_{\mu}$
is the electromagnetic field strength tensor. To obtain HBH solutions,
it showed that the dimensionless coupling constant $a$ has to be
larger than 1/4 \cite{Herdeiro:2018wub}. For the asymptotically flat
and spherically symmetric black hole solution ansatz, 
\begin{equation}
ds^{2}=-N(r)e^{-2\delta(r)}dt^{2}+\frac{dr^{2}}{N(r)}+r^{2}\left(d\theta^{2}+\sin^{2}\theta d\varphi^{2}\right)\text{,}\quad\mathbf{A}=A_{t}dt=V(r)dt,\label{eq:ansatz}
\end{equation}
the equations of motion are 
\begin{align}
2m^{\prime}(r)-r^{2}N(r)\phi^{\prime}(r)^{2}-e^{2\delta(r)+a\phi(r)^{2}}r^{2}V^{\prime}(r)^{2} & =0,\nonumber \\
\delta^{\prime}(r)+r\phi^{\prime}(r)^{2} & =0,\nonumber \\
\left[e^{-\delta(r)}r^{2}N(r)\phi^{\prime}(r)\right]^{\prime}-\alpha e^{\delta(r)+a\phi(r)^{2}}\phi(r)r^{2}V^{\prime}(r)^{2} & =0,\label{eq:EOM}\\
\left[e^{\delta(r)+a\phi(r)^{2}}r^{2}V^{\prime}(r)\right]^{\prime} & =0,\nonumber 
\end{align}
where the Misner-Sharp mass function $m(r)$ is defined by $N(r)\equiv1-2m(r)/r$,
and primes denote derivatives with respect to $r$. The last equation
in eqn. $\left(\ref{eq:EOM}\right)$ leads to $V^{\prime}(r)=-e^{-\delta(r)-a\phi(r)^{2}}Q/r^{2}$,
where the constant $Q$ can be interpreted as the electric charge
of the HBH. Moreover, one can implement boundary conditions at the
event horizon of radius $r_{h}$ and spatial infinity as 
\begin{align}
 & m\left(r_{h}\right)=\frac{r_{h}}{2}\text{, }\delta\left(r_{h}\right)=\delta_{0}\text{, }\phi\left(r_{h}\right)=\phi_{0}\text{, }V\left(r_{h}\right)=0\text{,}\nonumber \\
 & m(\infty)=M\text{, }\delta(\infty)=0\text{,}\quad\phi(\infty)=0\text{, }V(\infty)=\Psi\text{,}\label{eq:BC}
\end{align}
where $\delta_{0}$ and $\phi_{0}$ are two constants characterizing
the black hole solution, $M$ is the Arnowitt-Deser-Misner mass, and
$\Psi$ is the electrostatic potential. In addition, we focus on the
fundamental state of the HBH solution, where the scalar field $\phi(r)$
stays positive outside the event horizon. Note that $\phi_{0}=\delta_{0}=0$
correspond to RN black holes with $\phi=0$. Nevertheless, HBH solutions
with a non-trivial scalar field $\phi$ can exist for non-zero values
of $\phi_{0}$ and $\delta_{0}$. In this paper, a shooting method
built in the NDSolve function of Wolfram$\circledR$Mathematica is
employed to numerically solve eqn. $\left(\ref{eq:EOM}\right)$ with
the given boundary conditions $\left(\ref{eq:BC}\right)$. The metric
functions of two HBH solutions with $a=0.9$ and $M=1$ are presented
in FIG. \ref{fig:HBH}, where the\textbf{ }blue and orange lines denote
$Q=1.0541$ and $Q=1.0641$, respectively.

Light rays propagating in the HBH spacetime are described by the geodesic
equations 
\begin{equation}
\frac{d^{2}x^{\mu}}{d\lambda^{2}}+\Gamma_{\rho\sigma}^{\mu}\frac{dx^{\rho}}{d\lambda}\frac{dx^{\sigma}}{d\lambda}=0,\label{eq:geo}
\end{equation}
where $\lambda$ is the affine parameter, and $\Gamma_{\rho\sigma}^{\mu}$
is the Christoffel symbol. Using $ds^{2}=0$, one can rewrite the
radial component of the geodesic equations as 
\begin{equation}
e^{-2\delta(r)}\left(\frac{dr}{d\lambda}\right)^{2}=\frac{1}{b^{2}}-\frac{e^{-2\delta(r)}N(r)}{r^{2}},\label{eq:nullgeo}
\end{equation}
where we rescale the affine parameter $\lambda$ as $\lambda\rightarrow\lambda/|L|$,
$b=|L|/E$ is the impact parameter, and $L$ and $E$ are the conserved
angular momentum and energy of photons, respectively. From\textbf{
}eqn.\textbf{ }$\left(\ref{eq:nullgeo}\right)$, the effective potential
governing light rays is defined as 
\begin{equation}
V_{\text{eff}}(r)=\frac{e^{-2\delta(r)}N(r)}{r^{2}}.
\end{equation}
Unstable circular null geodesics at radius $r_{\text{ph}}$, which
constitute a photon sphere of radius $r_{\text{ph}}$, are determined
by 
\begin{equation}
V_{\text{eff}}\left(r_{\text{ph}}\right)=\frac{1}{b_{\text{ph}}^{2}},\quad V_{\text{eff}}^{\prime}\left(r\right)=0,\quad V_{\text{eff}}^{\prime\prime}\left(r_{\text{ph}}\right)<0,
\end{equation}
where $b_{\text{ph}}$ is the corresponding impact parameter. The
effective potential $V_{\text{eff}}(r)$ of the aforementioned HBH
solutions is plotted in the right panel of FIG. \ref{fig:HBH}. Remarkably,
it shows that, when the black hole charge is large enough, $V_{\text{eff}}(r)$
can have two local maxima, leading to a double-peak structure. Consequently,
hairy black holes with the double-peak effective potential possess
two photon spheres outside the event horizon.

In this paper, we study images of the HBHs illuminated by light sources
on a celestial sphere and use the numerical backward ray-tracing method
to calculate light rays from an observer to the celestial sphere (or
the event horizon). To supply initial conditions for eqn. $\left(\ref{eq:nullgeo}\right)$,
one considers the 4-momentum of a photon measured by a static observer
located at $\left(t_{\text{o}},r_{\text{o}},\theta_{\text{o}},0\right)$,
\begin{equation}
p^{(t)}=-\frac{p_{t}}{\sqrt{N(r_{\text{o}})}e^{-\delta(r_{\text{o}})}},\quad p^{(r)}=p_{r}\sqrt{N(r_{\text{o}})},\quad p^{(\theta)}=\frac{p_{\theta}}{r_{\text{o}}},\quad p^{(\varphi)}=\frac{p_{\varphi}}{r_{\text{o}}|\sin\theta_{\text{o}}|},
\end{equation}
where $p^{\mu}=\left.dx^{\mu}/d\lambda\right\vert _{\left(t_{\text{o}},r_{\text{o}},\theta_{\text{o}},0\right)}$.
As in \cite{Cunha:2016bpi}, the observation angles $\alpha$ and
$\beta$ are introduced as 
\begin{equation}
\sin\alpha=\frac{p^{(\theta)}}{p^{(t)}}\text{, }\tan\beta=\frac{p^{(\varphi)}}{p^{(r)}},
\end{equation}
which are determined by the momentum of the photon received by the
observer. The Cartesian coordinates $\left(x,y\right)$ of the image
plane of the observer is defined by 
\begin{equation}
x\equiv-r_{\text{o}}\beta,\text{ }y\equiv r_{\text{o}}\alpha,
\end{equation}
where the sign convention for $\beta$ leads to the minus sign in
the $x$ definition. Note that the zero observation angles $\left(0,0\right)$
correspond to the direction pointing to the center of the black hole.

The HBH image viewed by the observer is determined by gravitational
lensing, which maps the observation angles $\alpha$ and $\beta$
to a point in the celestial sphere or the event horizon. Lights rays
coming from the event horizon correspond to gray pixels in the HBH
image, which compose the black hole shadow. On the other hand, the
pattern information of the light source is retrieved by lights rays
connecting the celestial sphere with the observer. Specially, light
rays asymptotically originating from photon spheres form critical
curves in the HBH image, and hence photons captured near the critical
curves will have circled the HBH many times before reaching the observer.
As anticipated, higher-order images of the celestial sphere would
stack up near the critical curves.

Furthermore, to understand the pattern of the HBH image, we consider
multiple images of a point-like light source on the celestial sphere.
Due to gravitational lensing, light rays from the point-like light
source to the observer can go through different paths to create an
infinite series of images. Without loss of generality, we suppose
that the observer and the point-like light source are on the equatorial
plane and at $\left(t_{\text{o}},r_{\text{o}},\pi/2,0\right)$ and
$\left(t_{\text{s}},r_{\text{s}},\pi/2,\varphi_{\text{s}}\right)$,
respectively. In this case, the change of angular coordinate $\Delta\varphi$
of the light rays is given by 
\begin{equation}
\Delta\varphi=\left\{ \begin{array}{c}
\min\left(\varphi_{\text{s}},2\pi-\varphi_{\text{s}}\right)+n\pi\text{, }n=0,2,4\cdots\\
\max\left(\varphi_{\text{s}},2\pi-\varphi_{\text{s}}\right)+\left(n-1\right)\pi\text{, }n=1,3,5\cdots
\end{array}\right.,
\end{equation}
where the number $n$ can be used to label the multiple images. In
particular, $n=0$ corresponds to the primary image, $n=1$ to the
secondary image, and $n>1$ to the higher-order images.

\section{Numerical Results}

\label{sec:Numresult}

\begin{figure}[ptb]
\includegraphics[width=0.5\textwidth]{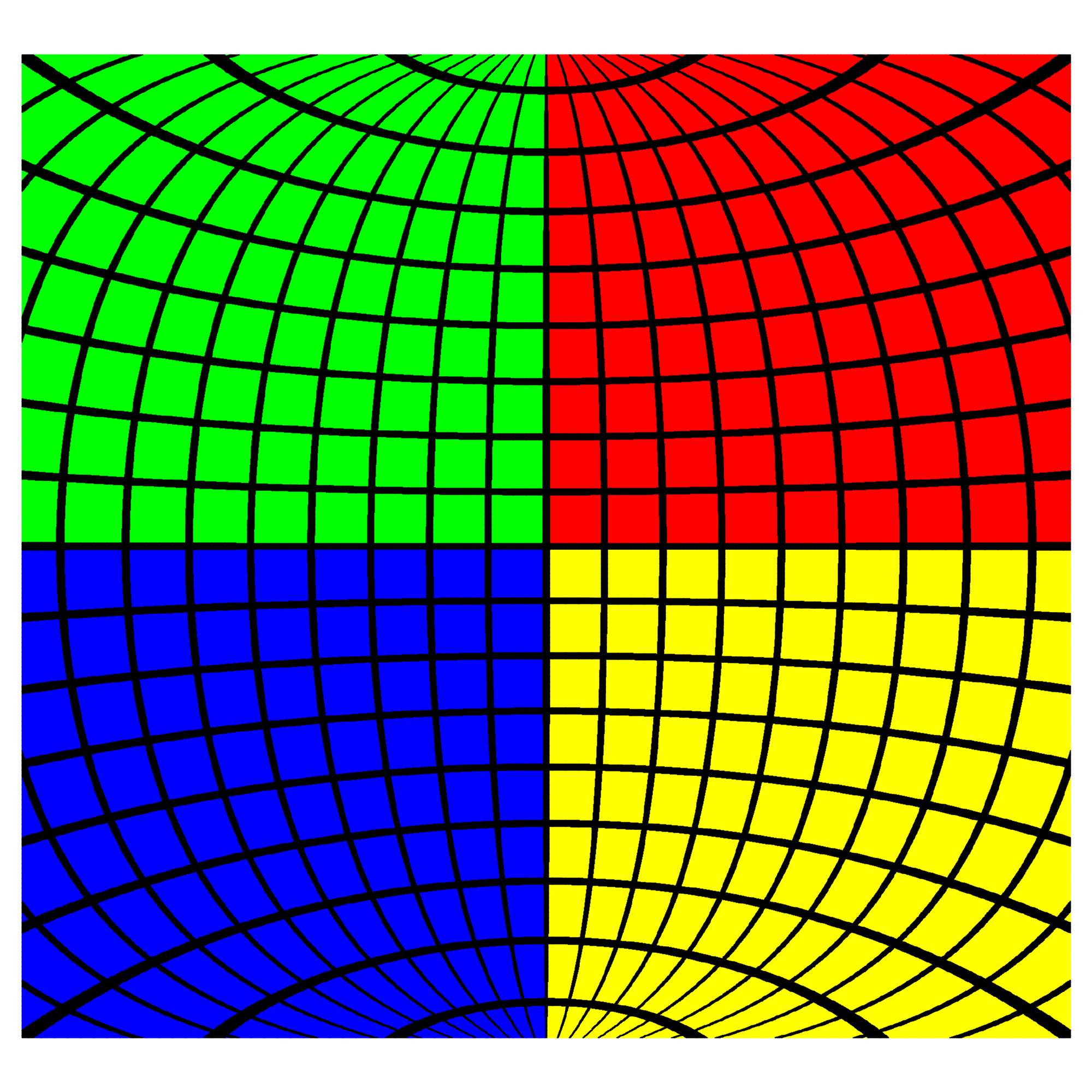}\caption{The observational image of the celestial sphere in the Minkowski spacetime.
The observer has a $2\pi/3$ field of view and is at $r_{O}=6$ on
the equatorial plane.}
\label{fig:flat}
\end{figure}

To illustrate gravitational lensing by HBHs with single-peak and double-peak
effective potentials, we place a luminous celestial sphere at $r_{\text{CS}}=12M$.
The celestial sphere is concentric with the HBHs and encloses observers.
Two static observers, henceforth denoted as $O$ and $P$, are located
at $x_{O}^{\mu}=\left(0,6M,\pi/2,0\right)$ and $x_{P}^{\mu}=\left(0,6M,\pi/3,0\right)$,
respectively, and their viewing angles capture $2\pi/3$ of the celestial
sphere. From the observers' position, we scan their viewing angles
by numerically integrating $3000\times3000$ null geodesics until
reaching the celestial sphere or hitting the event horizon. The light
source on the celestial sphere is divided into four quadrants, each
of which is painted with a different color. Moreover, we also lay
a set of black lines of constant longitude and latitude, with adjacent
lines separated by $\pi/180$. To illustrate the pattern of the light
source, we present the image of the celestial sphere viewed by the
observer $O$ in Minkowski spacetime in FIG. \ref{fig:flat}. Note
that an external view of the celestial sphere can be found in \cite{Bohn:2014xxa,Cunha:2015yba}.
Without loss of generality, we set $M=1$ in this section.

\subsection{Single-peak potential}

\label{subsec:Single-peak-potential}

\begin{figure}[t]
\includegraphics[scale=0.45]{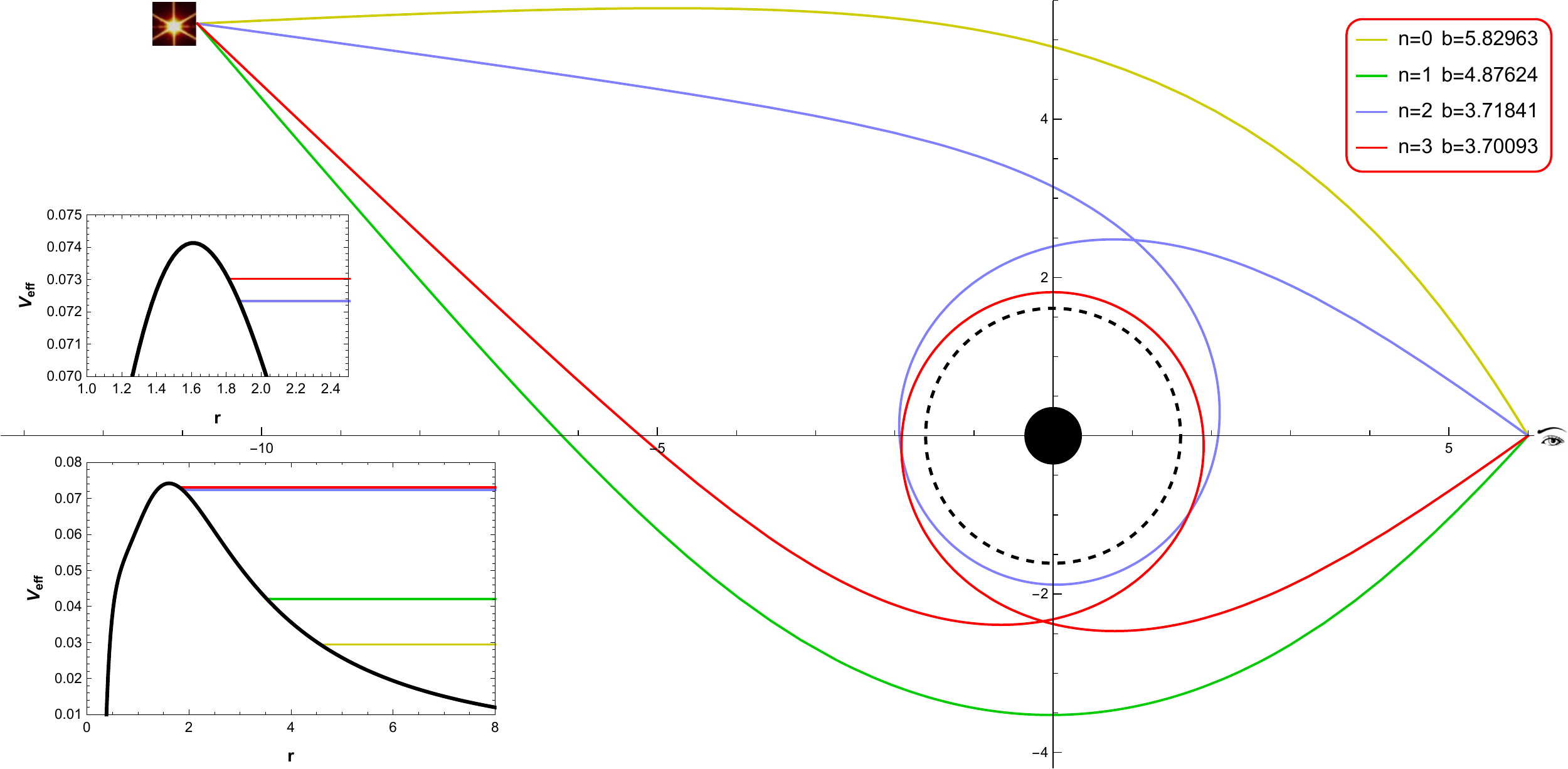}\caption{Light rays producing the primary ($n=0$, yellow line), secondary
($n=1$, green line) and higher-order ($n=2$ and $3$, blue and red
lines) images of a point-like light source on the equatorial plane
of the HBH with $a=0.9$ and $Q=1.0541$, which has a single-peak
potential. The black disk and dashed circle represent the HBH and
the photon sphere, respectively. The horizontal lines in the insets
denote $b^{-2}$ of the corresponding light rays, where $b$ is the
impact parameter. The impact parameter of the higher-order images
is very close to that of the photon sphere located at the potential
peak.}
\label{fig:raylines-single}
\end{figure}

We consider gravitational lensing by the HBH with the coupling $a=0.9$
and the electric charge $Q=1.0541$, which possesses a single-peak
effective potential as shown in the insets of FIG. \ref{fig:raylines-single}.
The single-peak potential indicates a single photon sphere, which
plays an important role in the gravitational lensing. In FIG. \ref{fig:raylines-single},
we plot four light rays connecting a point-like light source on the
equator of the celestial sphere with the observer $O$, which produce
the $n\leq3$ images of the light source seen by the observer $O$.
It displays that strong gravitational lensing near the photon sphere
is responsible for the higher-order images of the point-like light
source. In addition, $b^{-2}$ of the light rays are plotted as horizontal
lines in the insets of FIG. \ref{fig:raylines-single}, which shows
that the impact parameter of the light rays producing the higher-order
images approaches that of the photon sphere from above in the strong
deflection limit with $n\rightarrow\infty$. Hence, the higher-order
images of the light source lie outside and asymptotically approach
the critical curve associated with the photon sphere at the potential
peak.

\begin{figure}[ptb]
\begin{centering}
\includegraphics[scale=0.07]{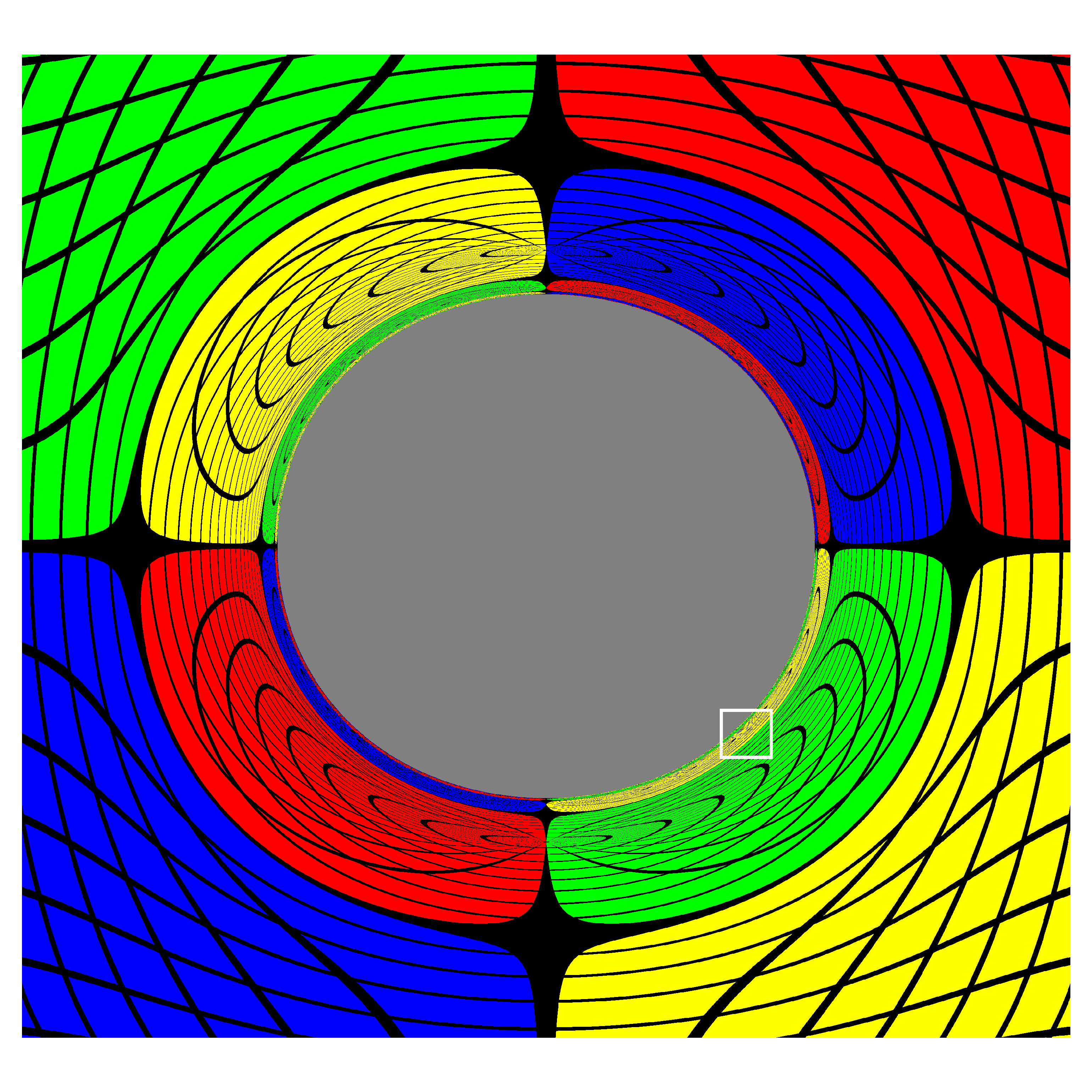} \includegraphics[scale=0.07]{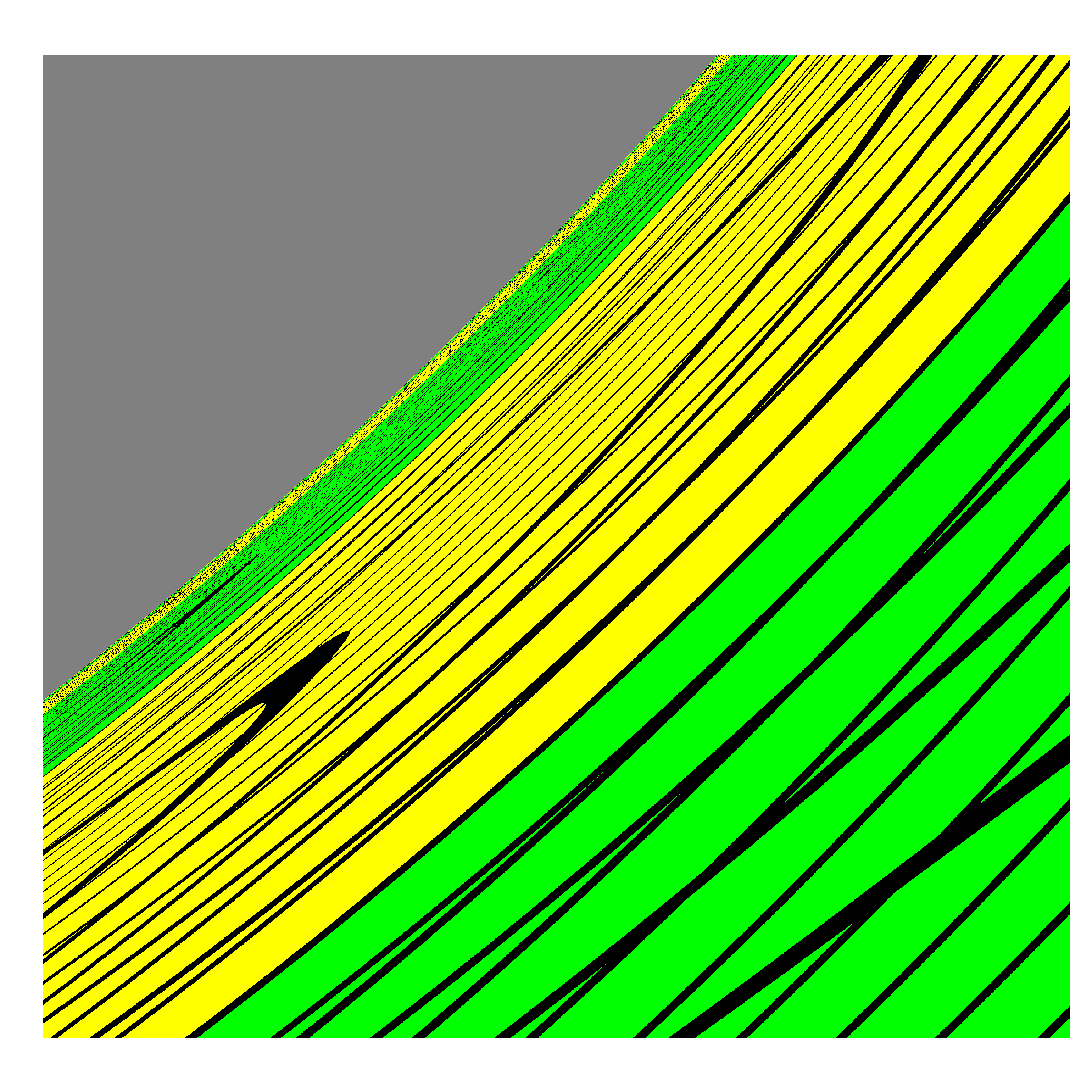}
\includegraphics[scale=0.07]{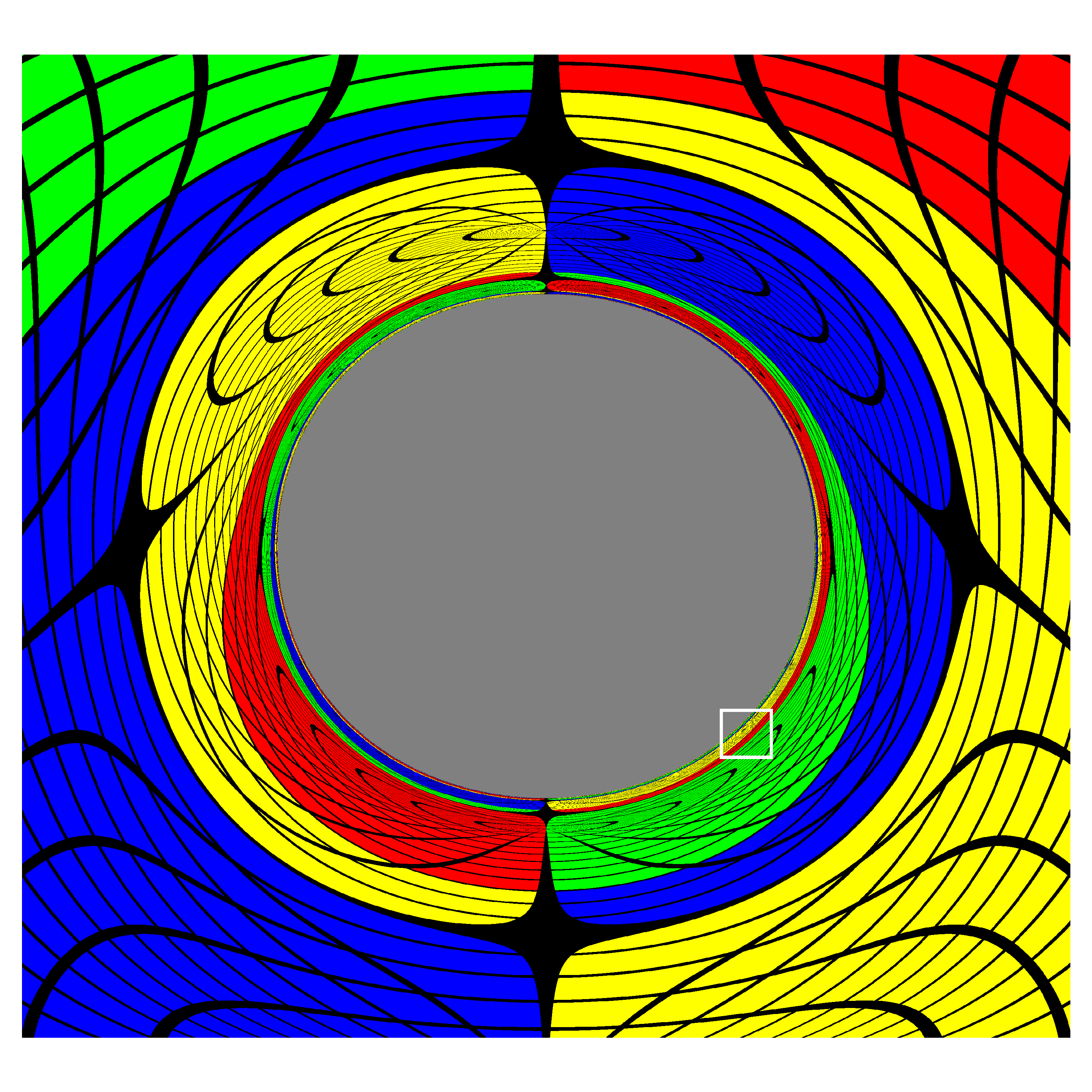} \includegraphics[scale=0.07]{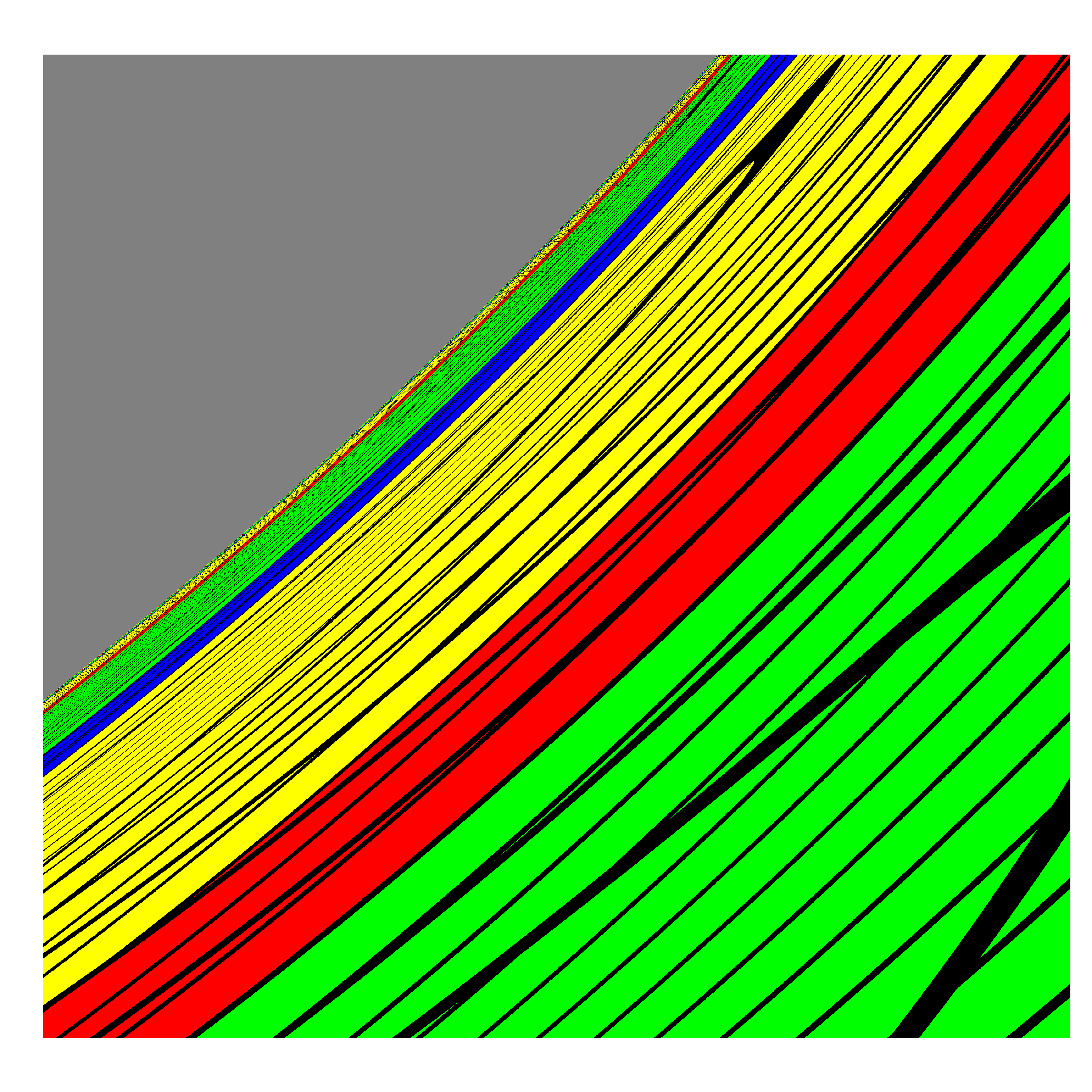} 
\par\end{centering}
\caption{Images of the HBH, which is illuminated by the celestial sphere, viewed
by the observers $O$ (top row) and $P$ (bottom row). Here, we take
$a=0.9$ and $Q=1.0541$, corresponding to a single-peak effective
potential. The panels in the right column highlight the regions near
the shadow edge (critical curve), which are bounded by white boxes
in the left column. There appears a series of higher-order images
of the celestial sphere, which asymptotically approaches the shadow
edge. }
\label{fig:singlepeak}
\end{figure}

When the HBH is illuminated by the celestial sphere, the black hole
images viewed by the observers $O$ and $P$ are presented in the
upper and lower rows, respectively, in FIG. \ref{fig:singlepeak}.
The gray area in the HBH images is the black hole shadow, whose edge
is the critical curve (or ``apparent boundary''\ called by Bardeen
\cite{Bardeen:1973tla}). As in the case of point-like light sources,
just outside the critical curve lies a sequence of higher-order celestial
sphere images, which forms the ``photon ring''\ \cite{Gralla:2019xty}.
To show detailed structure of the photon ring, we magnify the regions
bounded by white boxes of the left column in the right column, which
display that the higher-order celestial sphere images are self-similar
and asymptotically approach the shadow edge.

\subsection{Double-peak potential}

\label{subsec:Double-peak-potential}

\begin{figure}[t]
\includegraphics[scale=0.55]{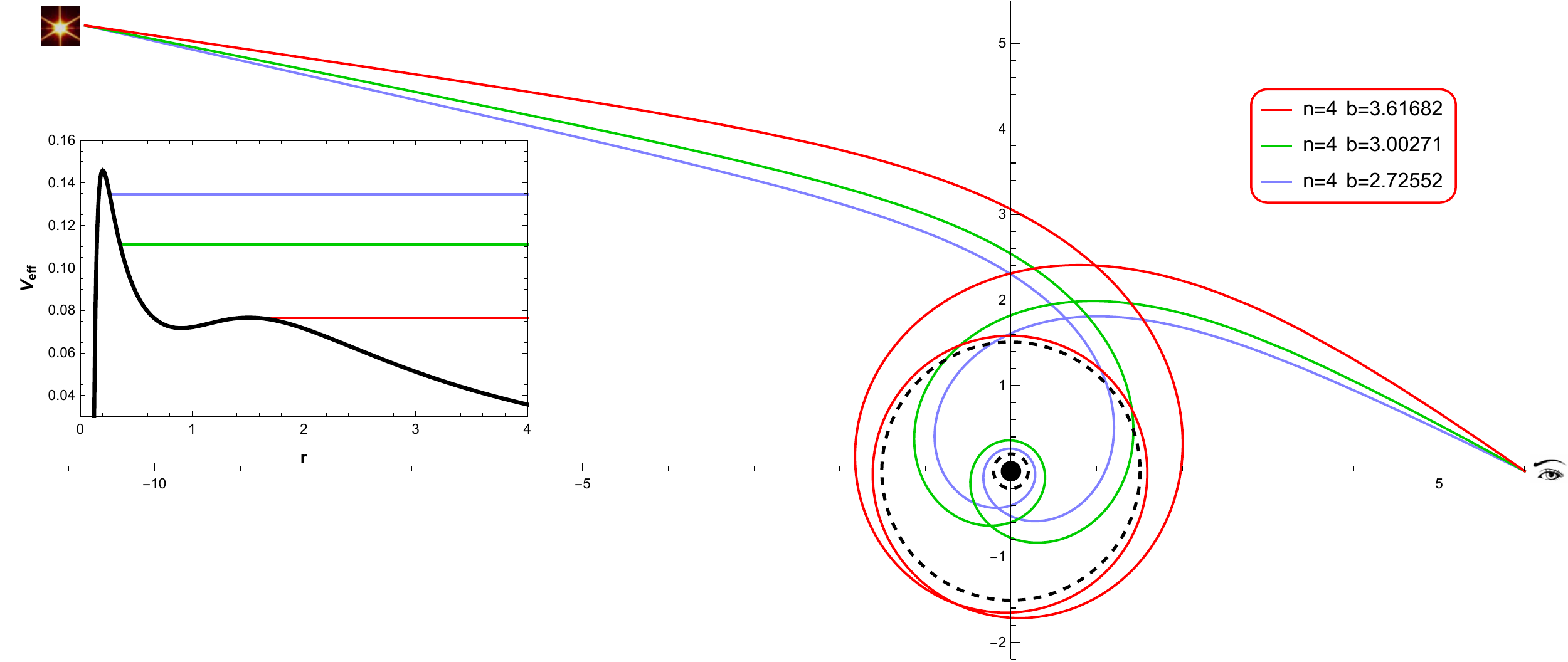} \includegraphics[scale=0.55]{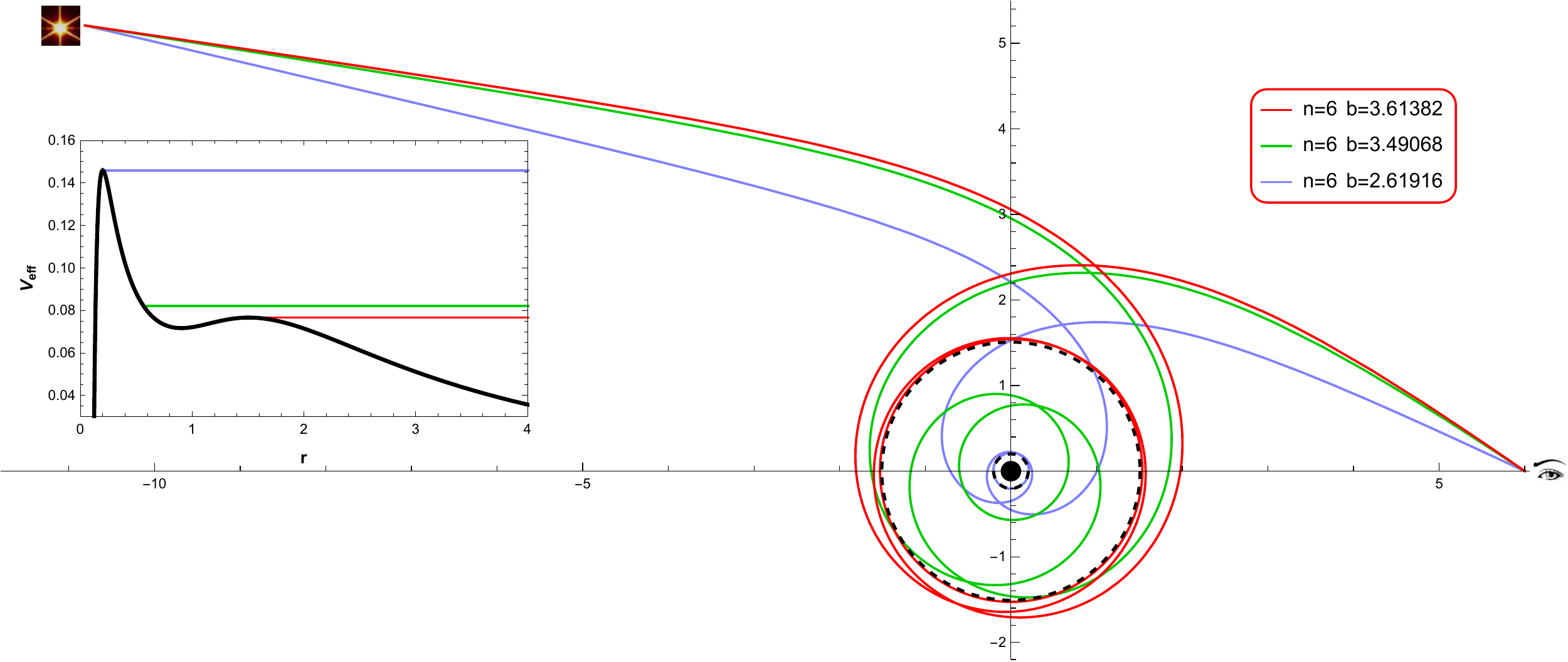}\caption{Light rays producing the $n=4$ (top panel) and $n=6$ (lower panel)
higher-order images of a point-like light source on the equatorial
plane of the HBH with $a=0.85$ and $Q=1.0603$, which has a double-peak
potential. The two dashed circles denote the photon spheres at the
potential peaks. In the insets, $b^{-2}$ of the light rays are plotted
as horizontal lines. For each $n$, strong gravitational lensing near
the photon spheres results in three higher-order images seen by the
observer $O$.\ }
\label{fig:raylines-double}
\end{figure}

We now investigate gravitational lensing by the HBH with the coupling
$a=0.85$ and the electric charge $Q=1.0603$, which possesses a double-peak
effective potential as shown in the insets of FIG. \ref{fig:raylines-double}.
The two peaks of the potential correspond to two photon spheres of
different size outside the event horizon. Note that the inner potential
peak is higher than the outer one, which makes effects of the smaller
photon sphere visible to a distant observer. Owing to the existence
of two photon spheres, higher-order images of some light source gravitationally
lensed by the HBH can have richer structure than in the single-peak
case.

FIG. \ref{fig:raylines-double} exhibits light rays travelling from
a point-like light source on the equator of the celestial sphere to
the observer $O$, which produce $n=4$ and $n=6$ higher-order images
of the light source, in the upper and lower panels, respectively.
What stands out in this figure is that, for a given $n$, three higher-order
images can be seen by the observer. On the other hand, a point-like
light source only produces one higher-order image with a fixed $n$
in a HBH of single-peak potential. To gain an insight into the higher-order
images, we present $b^{-2}$ of the light rays as horizontal lines
in the insets of FIG. \ref{fig:raylines-double}.\ The impact parameter
$b$ of the light rays with the largest/smallest $b$, which are displayed
as red/blue lines, is slightly larger than $b$ of the larger/smaller
photon sphere, indicating that the light rays of largest/smallest
$b$ may circle around the larger/smaller photon sphere several times
before captured by the observer, and produce higher-order images outside
the critical curve associated with the larger/smaller photon sphere.
When $n$ is large, the light rays with the largest and smallest $b$
are expected to be temporarily trapped around the larger and smaller
photon spheres, respectively. More interestingly, the light rays of
intermediate $b$, which are represented by green lines, can circle
around the HBH between the larger and smaller photon spheres. As shown
in the insets of FIG. \ref{fig:raylines-double}, the impact parameter
$b$ of the light rays with the intermediate $b$ is smaller than
that of the larger photon sphere and approaches it with increasing
$n$. This observation suggests that, these light rays generate higher-order
images inside the critical curve associated with the larger photon
sphere, and would circle many times around the larger photon sphere
in the strong deflection limit when they travel toward or away from
the HBH. Note that light rays with $b$ asymptotically smaller than
that of a photon sphere have been analytically discussed in the framework
of ultracompact objects \cite{Shaikh:2019itn} and wormholes \cite{Tsukamoto:2021caq}.

\begin{figure}[ptb]
\begin{centering}
\includegraphics[scale=0.07]{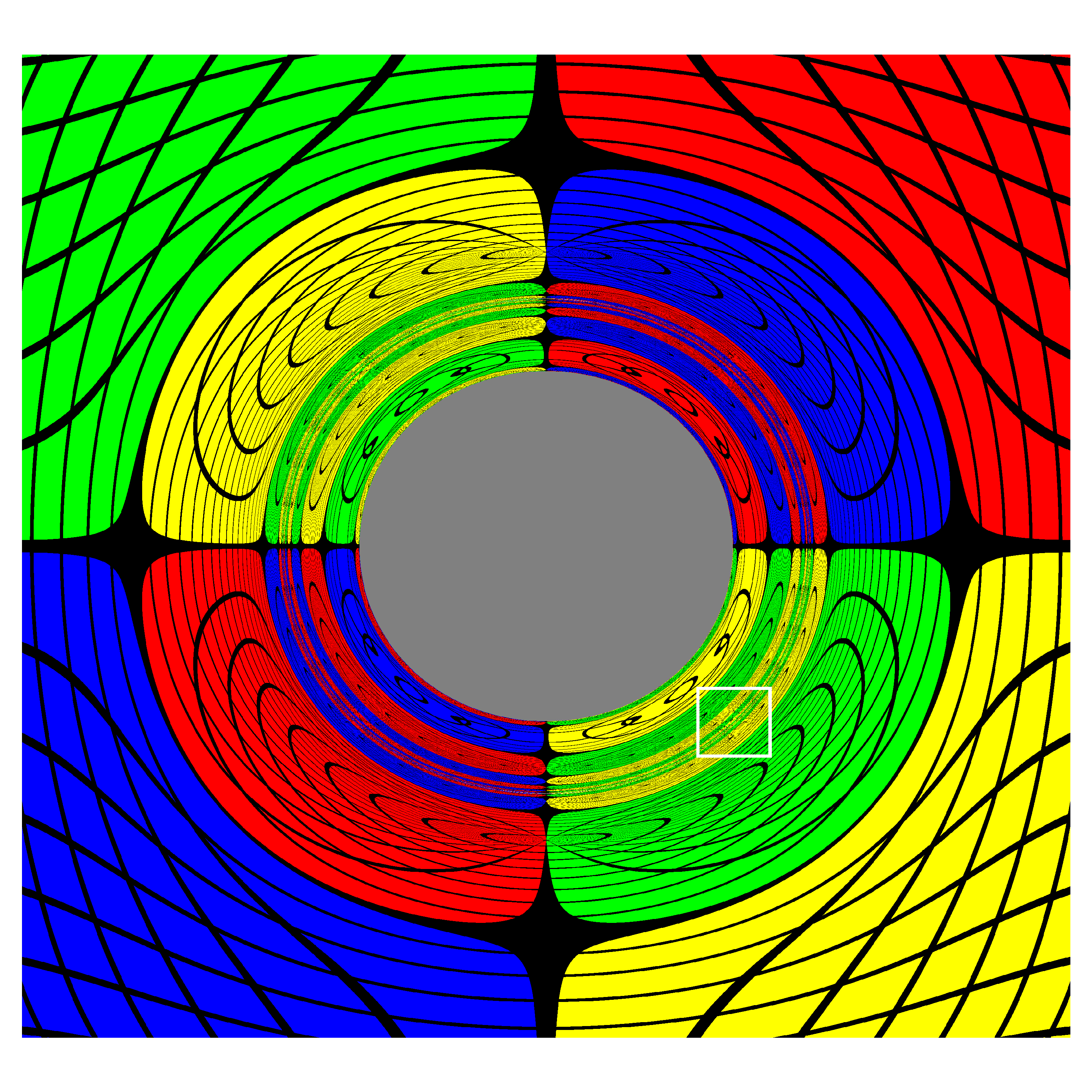} \includegraphics[scale=0.07]{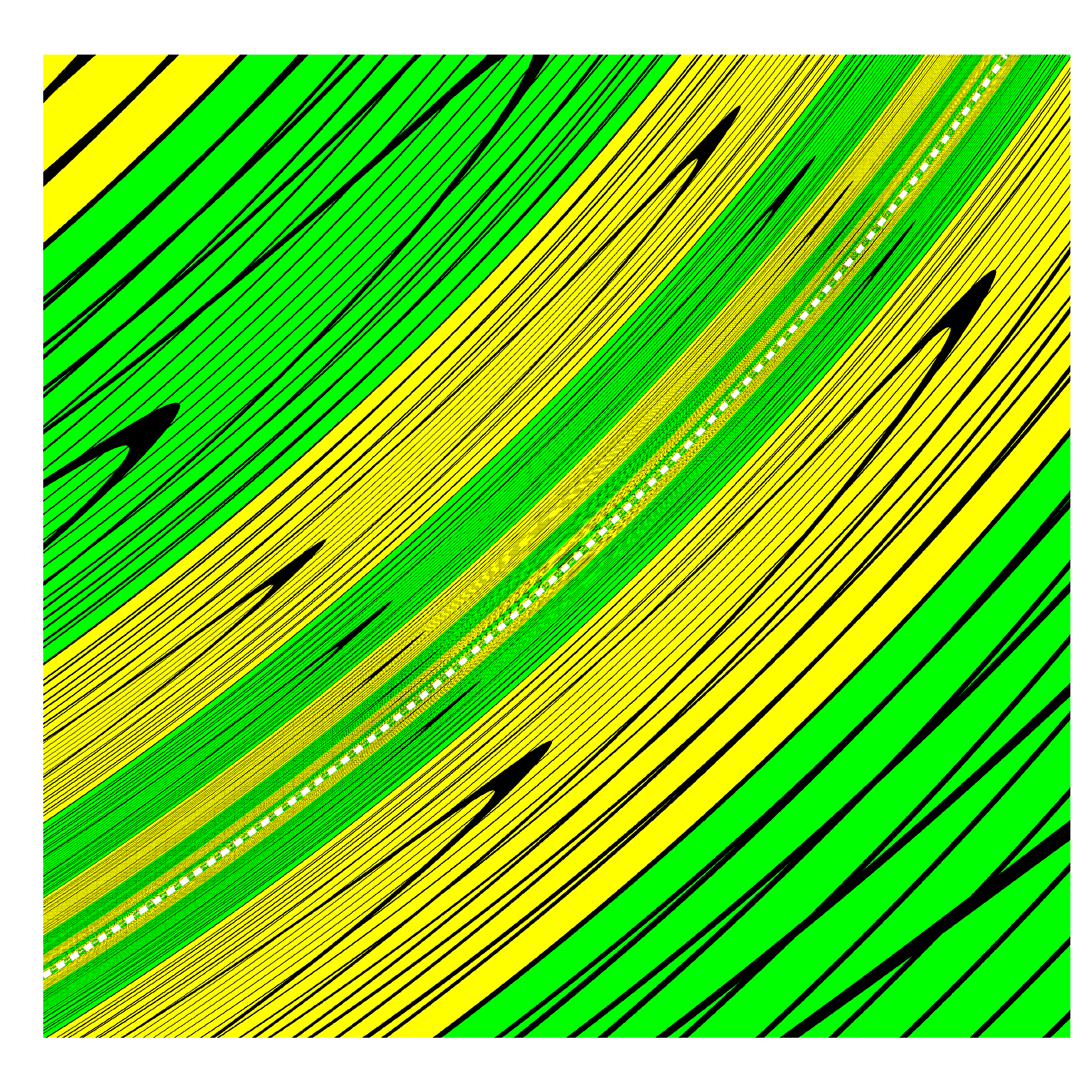}
\includegraphics[scale=0.07]{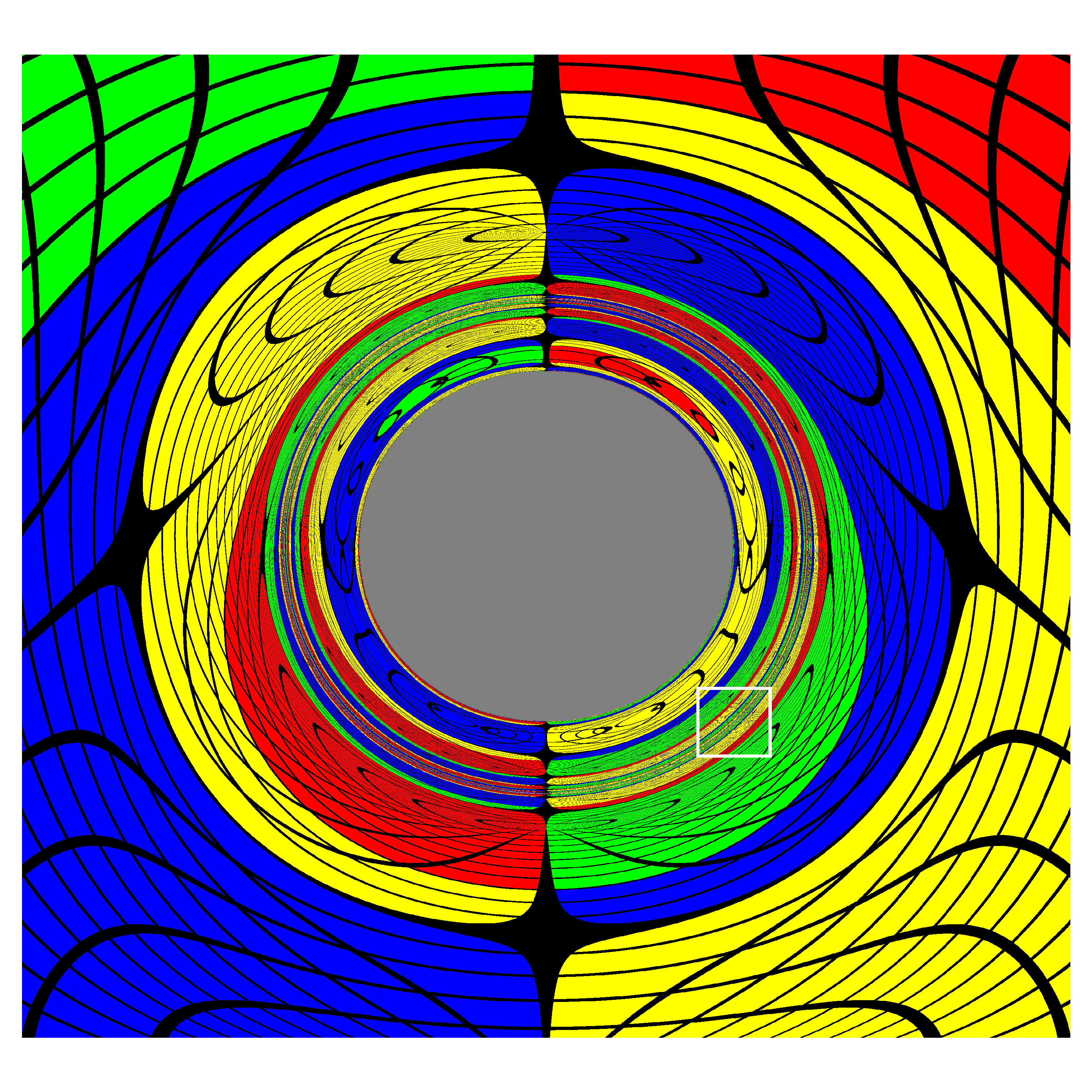} \includegraphics[scale=0.07]{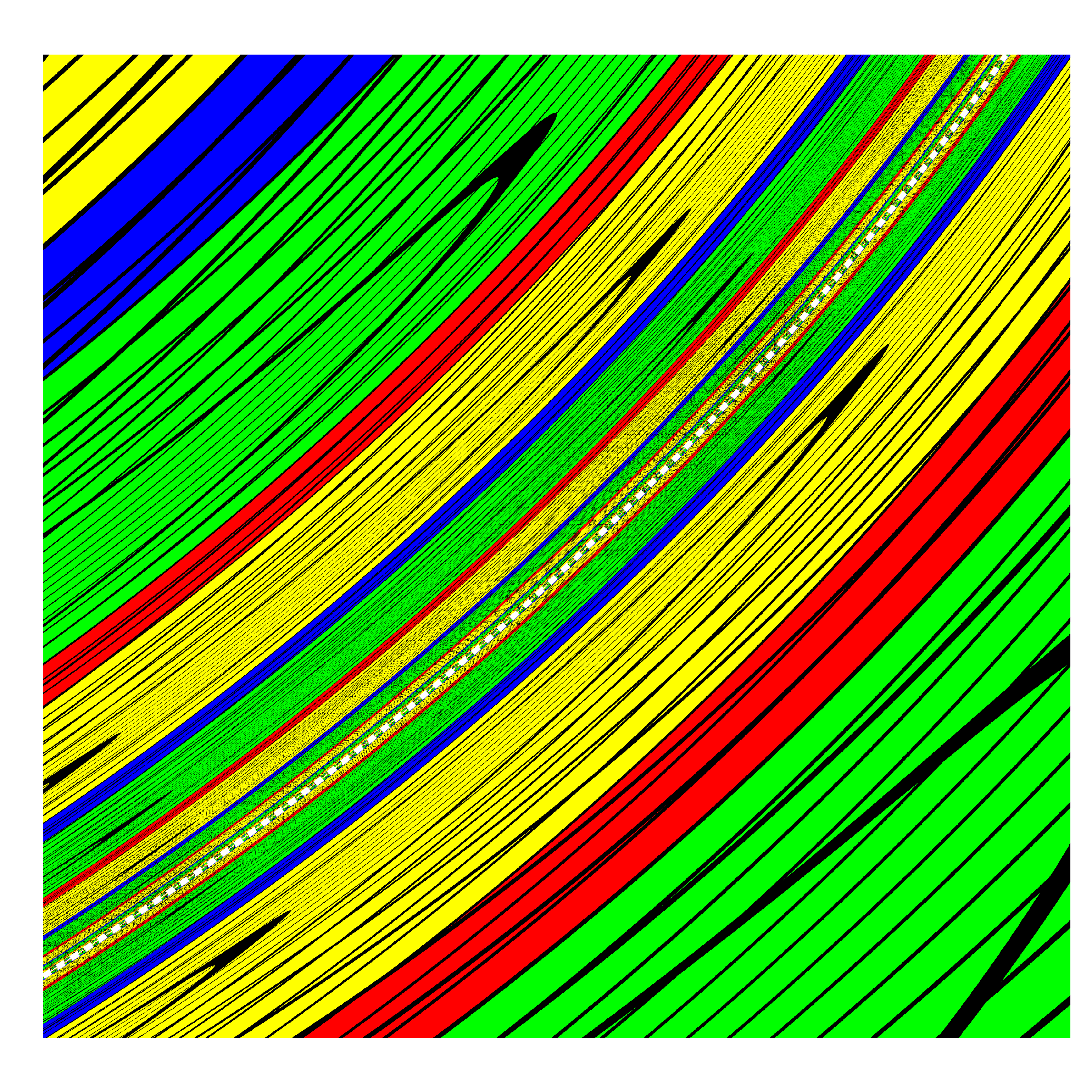} 
\par\end{centering}
\caption{Images of the celestial sphere gravitationally lensed by the HBH with
$a=0.85$ and $Q=1.0603$, which has a double-peak potential. The
observers $O$ and $P$ are considered in the upper and lower rows,
respectively. The panels in the right column zoom in the regions bounded
by white boxes in the left column. The shadow edge and the dashed
white line are the critical curves, which are determined by the smaller
and larger photon spheres, respectively. Besides a set of higher-order
images just outside the shadow edge, two sets of higher-order images
can be seen near the other critical curve.}
\label{fig:doulepeak}
\end{figure}

If the light source is the celestial sphere, strong gravitational
lensing near two photon spheres can leave distinctive imprints on
the HBH images. Specifically, light rays asymptotically approaching
the two photon spheres yield two critical curves in the HBH images.
As explained earlier, higher-order images of the celestial sphere
are stacked up near the critical curves. Moreover, the results from
the previous paragraph signal that three sets of higher-order images
can be observed in the HBH images. In fact, images of the HBH illuminated
by the celestial sphere are displayed for the observers $O$ and $P$
in the upper and lower rows of FIG. \ref{fig:doulepeak}, respectively.
Similar to the single-peak case, the shadow edge is one critical curve
determined by the smaller photon sphere, and one set of higher-order
images of the celestial sphere lies just outside the shadow edge.
The zoom-in images of the regions bounded by white boxes in the left
column are exhibited in the right column, where dashed white lines
represent the other critical curve determined by the larger photon
sphere. Remarkably, two more sets of higher-order images are spotted
inside and outside this critical curve, respectively. In the strong
deflection limit, these higher-order images asymptotically approach
the critical curve. In contrast to the single-peak case, the photon
ring of the images of HBHs with a double-peak potential includes three
sets of higher-order images.

\section{Conclusions}

\label{sec:CON}

In this paper, we investigated lensed images of a point-like light
source and a\ luminous celestial sphere for a class of HBHs in an
Einstein-Maxwell-scalar theory, where the scalar field is minimally
coupled to the electromagnetic sector. Depending on the HBH parameters,
the HBHs can have one or two photon spheres outside the event horizon.
For the HBHs with a single photon sphere, strong gravitational lensing
around the photon sphere brings about higher-order images of the point-like
light source and a set of higher-order images of the celestial sphere
just outside the shadow edge. On the other hand, the existence of
an extra photon sphere could triple the number of the higher-order
images of the point-like light source. Moreover, three sets of higher-order
images of the celestial sphere near two critical curves are observed
for the HBHs with two photon spheres. These peculiar signatures provide
a powerful tool to search for black holes with multiple photon spheres.

While the shadow size of a black hole may depend on the surrounding
astrophysical environment \cite{Gralla:2019xty,Chael:2021rjo}, the
photon rings, which are composed of higher-order images of the astrophysical
source, have been found to be insensitive to the assumed astrophysical
source model \cite{Gralla:2020srx}. Consequently, detections of the
photon rings would probe the underlying spacetime in a model-independent
way, providing powerful new tests of general relativity and Kerr hypothesis
\cite{Johnson:2019ljv,Hadar:2020fda,Wielgus:2021peu,Broderick:2021ohx}.
To explore the effects of multiple photon spheres on future detections
of the photon rings, it will be of great interest if our analysis
can be generalized to more astrophysically realistic models.
\begin{acknowledgments}
We are grateful to Yiqian Chen for useful discussions and valuable
comments. This work is supported in part by NSFC (Grant No. 12105191,
11947225 and 11875196). Houwen Wu is supported by the International
Visiting Program for Excellent Young Scholars of Sichuan University. 
\end{acknowledgments}

 \bibliographystyle{unsrturl}
\bibliography{ref}

\end{document}